\documentclass[a4paper,11pt]{article}
\pdfoutput=1 

\usepackage{jheppub} 
\usepackage[T1]{fontenc} 
\usepackage{amsmath,amssymb}
\usepackage{dsfont}
\usepackage{upgreek}
\usepackage{slashed}
\usepackage{appendix}
\usepackage{tabu}
\usepackage{physics}
\usepackage{xcolor}

\usepackage{bbm}
\usepackage{mathrsfs}
\usepackage{multirow}
\usepackage{relsize}
\usepackage{caption}
\usepackage{cancel}
\usepackage{graphics,graphicx}
\usepackage{float}
\usepackage{lipsum}
\usepackage{comment}
\usepackage{enumitem}
\numberwithin{equation}{section}
\def\beq{\begin{align}}
\def\eeq{\end{align}}
\newcommand{\bi}{\begin{itemize}}
\newcommand{\ei}{\end{itemize}}
\newcommand{\ben}{\begin{enumerate}}
\newcommand{\een}{\end{enumerate}}
\newcommand{\be}{\begin{equation}}
\newcommand{\ee}{\end{equation}}
\newcommand{\bea}{\begin{eqnarray}}
\newcommand{\eea}{\end{eqnarray}}
\def\SM{{\scriptscriptstyle SM}}
\def\H{{\scriptscriptstyle H}}
\usepackage{tikz}
\usepackage{subcaption}

\title{Starobinsky Inflation from String Theory?}
\author[a,b,c]{Max Brinkmann,}
\author[c,d]{Michele Cicoli,}
\author[c,e,f]{Pietro Zito}

\affiliation[a]{\footnotesize Dipartimento di Fisica e Astronomia, Universit\`a di Padova, via Marzolo 8, 35131 Padova, Italy}
\affiliation[b]{\footnotesize INFN, Sezione di Padova, via Marzolo 8, 35131 Padova, Italy}
\affiliation[c]{\footnotesize Dipartimento di Fisica e Astronomia, Universit\`a di Bologna, via Irnerio 46, 40126 Bologna, Italy}
\affiliation[d]{\footnotesize INFN, Sezione di Bologna, viale Berti Pichat 6/2, 40127 Bologna, Italy}
\affiliation[e]{\footnotesize Institute for Quantum Optics and Quantum Information (IQOQI) Vienna, Austrian Academy of Sciences, Boltzmanngasse 3, 1090 Vienna, Austria}
\affiliation[f]{\footnotesize University of Vienna, Faculty of Physics \& Vienna Doctoral School in Physics, Boltzmanngasse 5, A-1090 Vienna, Austria}

\emailAdd{brinkmann@pd.infn.it}
\emailAdd{michele.cicoli@unibo.it}
\emailAdd{pietro.zito@oeaw.ac.at}

\abstract{Starobinsky inflation is currently one of the best models concerning agreement with cosmological data. Despite this observational success, it is still lacking a robust embedding into a UV complete theory. Previous efforts to derive Starobinsky inflation from string theory have been based on the derivation of higher derivative curvature terms from the low-energy limit of ten-dimensional string theory. This approach is however known to fail due to the difficulty to tame the effect of contributions proportional to the Ricci scalar to a power larger than two. In this paper we investigate an alternative attempt which exploits instead the ubiquitous presence of scalar fields in string compactifications combined with the fact that Starobinsky inflation can be recast as Einstein gravity coupled to a scalar field with a precise potential and conformal coupling to matter fermions. We focus in particular on type IIB K\"ahler moduli since they have shown to lead to exponential potentials with a Starobinsky-like plateau. We consider three classes of moduli with a different topological origin: the volume modulus, bulk fibre moduli, and blow-up modes. The only modulus with the correct coupling to matter is the volume mode but its potential does not feature any plateau at large field values. Fibre moduli admit instead a potential very similar to Starobinsky inflation with a natural suppression of higher curvature corrections, but they cannot reproduce the correct conformal coupling to matter. Blow-up modes have both a wrong potential and a wrong coupling. Our analysis implies therefore that embedding Starobinsky inflation into string theory seems rather hard. Finally, it provides a detailed derivation of the coupling to matter of fibre moduli which could be used as a way to discriminate Starobinsky from fibre inflation.}

\begin{document} 
\maketitle
\flushbottom

\section{Introduction}
\label{Intro}

Starobinsky inflation \cite{Starobinsky:1980te} is presently one of the most successful inflationary models in fitting cosmological observations \cite{Planck:2018jri}. This predictive model is based on a simple extension of the Einstein-Hilbert action via the introduction of an $R^2$ contribution. Applying a conformal transformation on the Jordan frame metric, the model can be recast as ordinary gravity in Einstein frame coupled to a scalar field $\phi$ with an exponential potential which asymptotes to an inflationary slow-roll plateau at large field values. Moreover, the conformal transformation of the metric fixes the Yukawa coupling of the inflaton to ordinary matter to be $y_\phi=-1/\sqrt{6}$. While elegant, this formulation of Starobinsky inflation is still lacking a quantum gravity embedding which is crucial to trust its robustness, especially against higher derivative curvature corrections which are naively expected to arise from the effective field theory point of view and which ruin the flatness of the inflationary potential. In fact, a working UV embedding of the Starobinsky model should explain why corrections to the Einstein-Hilbert action involving the Ricci and Riemann tensors are absent or can be ignored. Moreover, it should be characterised by at least two mass scales: $M\simeq 10^{13}$ GeV, which controls the $R^2$ contribution, and a much larger scale, $M_*\gg M$, which suppresses higher curvature terms $R^n$ with $n>2$.

String theory is at the moment one of best developed candidate theories of quantum gravity, and so it is natural to try to embed Starobinsky inflation in this framework.  In string theory higher derivative corrections to the four-dimensional Einstein-Hilbert action do in fact arise as the low-energy limit of $\alpha'$ corrections in ten dimensions. They are naturally suppressed by the string scale with potentially additional powers of the vacuum expectation values of universal moduli like the internal volume or the dilaton which fixes the string coupling. This observation gives some hope to obtain more than one suppression scale to justify the inclusion of just $R^2$ effects. However, as argued in \cite{Burgess:2016owb}, a detailed analysis of the low-energy limit of both heterotic and type II string theories reveals that letting the first higher derivative term compete with the leading term results in a loss of control over the perturbative $\alpha'$ expansion. In summary, allowing higher order terms to scale as expected from string theory quickly destroys any inflationary dynamics.

In this paper we will instead explore a different attempt to derive Starobinksy inflation from string theory based on the dual formulation of the model in terms of Einstein gravity coupled to a scalar field, the inflaton. Since string theory features many scalar fields, called moduli, arising as Kaluza-Klein zero modes of deformations of the internal compactification space, our hope is to reinterpret one of these scalars as the Starobinsky inflaton and recover the Starobinsky model in Einstein frame. This modulus should have two important features: ($i$) a scalar potential which reproduces the one of Starobinsky inflation, with in particular a plateau at large field values which can sustain at least $50$-$60$ efoldings of slow-roll inflation without being destroyed by higher order corrections (like $R^n$ terms with $n>2$ in Jordan frame); ($ii$) the correct conformal Yukawa coupling to matter coming from minimally coupled fermions in Jordan frame.

Several different mechanisms to drive inflation from string theory have been derived, both at single and a multi-field level (see \cite{Cicoli:2023opf} for a recent and updated review). Focusing on single-field models (as in the Starobinsky case), the most popular inflaton candidates are open string moduli, axions and K\"ahler moduli. The potential of open string moduli and axions is typically power-law, or at most sinusoidal for axions, while K\"ahler moduli feature exponential potentials \cite{Cicoli:2011zz,Burgess:2014tja} which arise naturally in the supergravity effective field theory of string compactifications once the potential is expressed in terms of canonically normalised fields. Given that the potential of the Starobinsky model is characterised by an exponential dependence on the inflaton, we will focus on K\"ahler moduli within type IIB compactification where moduli fixing is better understood. The stabilisation of these modes is crucial, not just to compute the inflationary potential, but also to derive their couplings to fermions living on stacks of either D7-branes wrapping internal four-cycles or D3-branes at singularities. 

We will focus on three different classes of K\"ahler moduli in type IIB Calabi-Yau compactifications. From the microscopic point of view, they are different since they control the volume of internal four-cycles with a different topology. Let us summarise our results for each of these classes of K\"ahler moduli:
\begin{itemize}
\item \textbf{Volume mode:} The volume mode has the correct conformal Yukawa coupling to matter fermions living on either D3-branes or D7-branes wrapped around a blow-up mode. It also has the right coupling to open string modes living within the worldvolume of D7-branes wrapping the overall volume. However, the potential of the volume mode cannot mimic the one of Starobinsky inflation since it does not feature any plateau at large field values. On the contrary, the potential of the volume modulus has a steep runaway at large field values and realisations of volume mode inflation require to tune different terms to induce a near inflection point \cite{Conlon:2008cj, Cicoli:2015wja, Antoniadis:2020stf}.

\item \textbf{Fibre moduli:} Fibre inflation is a class of models where inflation is driven by a bulk fibre modulus which is a leading order flat direction lifted by subdominant perturbative corrections to the K\"ahler potential \cite{Cicoli:2008gp, Broy:2015zba, Cicoli:2016chb, Cicoli:2016xae, Cicoli:2017axo, Cicoli:2020bao, Bhattacharya:2020gnk}. These effects generate a potential that is very similar to Starobinsky inflation. In fact, when recast as an $f(R)$ theory, fibre inflation looks effectively like $f(R)\simeq R^{1.3} + R^2$ \cite{Broy:2014xwa}. Its predictions for the main cosmological observables are thus very similar to the predictions of Starobinsky inflation. If the relation between the scalar spectral index $n_s$ and the tensor-to-scalar ratio $r$ is written as $r=3\alpha (n_s-1)^2$, Starobinsky inflation corresponds to $\alpha=1$, while fibre inflation to $\alpha=2$ \cite{Burgess:2016owb}. However, fibre moduli can never reproduce the correct Yukawa coupling to matter since they are effectively decoupled from fermions on D3-branes or on D7-branes wrapping blow-up modes. On the other hand, they have a non-zero coupling to fermions if the Standard Model is realised on branes wrapping bulk four-cycles, but the actual value of the Yukawa coupling is always different from $y_\phi =-1/\sqrt{6}$. More precisely, matching this value would require irrational modular weights in the K\"ahler metric for matter fields, a situation which seems impossible to achieve.

\item \textbf{Blow-up modes:} Moduli controlling the volume of local exceptional divisors resolving point-like singularities seem the worst candidates to reproduce Starobinsky inflation since both their potential and coupling take the wrong form. Despite featuring an exponential dependence, the potential of blow-up inflation is much flatter than the one of Starobinsky inflation \cite{Conlon:2005jm, Bond:2006nc, Blanco-Pillado:2009dmu}. Moreover, blow-up modes are effectively decoupled from fermions on D3-branes or on D7-branes wrapping bulk cycles, while they have a much stronger than Planckian coupling to matter on D7-branes wrapped around the blow-up mode itself (due to the localisation of the interaction in the extra dimensions).
\end{itemize}

As a result of our analysis, we conclude that no string modulus seems to have the right properties to effectively realise an $R+R^2$ inflationary model, at least within the framework of type IIB Calabi-Yau compactifications at large volume and weak string coupling where the low-energy effective field theory is under control. Embedding Starobinsky inflation in a UV complete theory like string theory seems therefore to be very challenging. If agreement with cosmological observations will become even better after the inclusion of more data, our analysis suggests that the underlying model is more likely to be fibre inflation which can be discriminated from Starobinsky inflation, not just from the slightly different relation between $n_s$ and $r$, but also from the different coupling to matter fermions. For example, we find that for matter at the intersection of D7-branes wrapping bulk four-cycles, the Yukawa coupling of fibre moduli is $y_\phi= 1/\sqrt{3}$. 

\section{Basics of Starobinsky inflation}
\label{sec:Starobinsky}

Starobinsky inflation \cite{Starobinsky:1980te} uses higher derivative corrections to the Einstein–Hilbert action, in particular the $R^2$ correction. Let us first review how this theory is classically equivalent to a standard Einstein gravity coupled to a scalar field with a scalar potential characterised by a large field plateau that can drive inflation. We shall then extend the scenario to more general $f(R)$ theories.

\subsection{$R+R^2$ gravity}

The starting point of Starobinsky inflation \cite{Starobinsky:1980te} is the following action
\begin{equation}
S=\frac12\,M_P^2\int\mathrm{d}^{4}x\sqrt{-g}\,f(R) 
+\int\mathrm{d}^{4}x\,{\cal L}_{\rm mat}(g_{\mu\nu},\psi)
\,,
\end{equation}
where
\begin{equation}
f(R)=R+\frac{R^{2}}{M^2}\,,
\label{eq:f(R)}
\end{equation}
and $\mathcal{L}_{\rm mat}$ describes the minimal coupling to matter fields collectively denoted as $\psi$. Here $M_P=1/\sqrt{8\pi G}$ and $M$ is a constant with the dimension of a mass. The Einstein-Hilbert linear term, $R$, is responsible for a deviation from an eternal de Sitter solution: it will cause inflation to end as required in any realistic inflationary model. 
However, the conformally equivalent form of the model is more useful for our purposes. In order to understand the universal features of the inflaton model, let us describe the conformal transformation in some detail.

A conformal transformation is a point-dependent rescaling of the metric tensor $g_{\mu\nu}$ of the form
\begin{equation}
g_{\mu\nu}\rightarrow\Tilde{g}_{\mu\nu}=\Omega(x)^2 g_{\mu\nu} 
    = e^{2\,\omega(x)} g_{\mu\nu}\,,
\end{equation}
with the scale factor $\omega(x)\equiv\ln \Omega(x)$.
One of its properties is to leave the light cones unchanged \cite{Faraoni:1998qx}.
The Ricci tensor $R_{\mu\nu}$ and the Ricci scalar $R$ constructed out of the metric $g_{\mu\nu}$, 
and $\Tilde{R}_{\mu\nu}$ and $\Tilde{R}$ obtained from $\Tilde{g}_{\mu\nu}$, 
are related by \cite{DeFelice:2010aj}
\begin{equation}
\label{eq:Rtrafo}
    R=\Omega^{2}(\tilde{R}+6\tilde{\square}\omega-6\tilde{g}^{\mu\nu}\partial_{\mu}\omega\partial_{\nu}\omega)\,,
\end{equation}
where 
\begin{equation}
    \partial_{\mu}\omega\equiv\frac{\partial\omega}{\partial x^{\mu}}\,,\qquad\tilde{\square}\omega\equiv\frac{1}{\sqrt{-
\tilde{g}}}\partial_{\mu}\left(\sqrt{-\tilde{g}}\,\tilde{g}^{\mu\nu}\partial_{\nu}
\omega\right)\,.
\end{equation}

We want to use this transformation to derive an action in Einstein frame from the Starobinsky model.
As a preliminary step, it is useful to rewrite the action in the equivalent form \cite{DeFelice:2010aj}
\begin{equation}
    S=\int\mathrm{d}^{4}x\sqrt{-g}\left(\frac12\,M_P^2 FR-U\right)+\int
\mathrm{d}^{4}x\,{\cal L}_{\rm mat}(g_{\mu\nu},\psi)\,,
\end{equation}
where we defined
\begin{equation}
     F\equiv \frac{\partial f}{\partial R} = 1+2\frac{R}{M^2}\;,
     \qquad U \equiv \frac12\,M_P^2\left(FR-f\right) = \frac12\,M_P^2\,\frac{R^2}{M^2}\,.
\end{equation}
Using the relation $\sqrt{-g}=\Omega^{-4}\sqrt{-\Tilde{g}}$ and \eqref{eq:Rtrafo}, the action becomes
\begin{equation}
S=\int\mathrm{d}^4x\,\sqrt{-\tilde{g}}\left[\frac12\,M_P^2 \frac{F}{\Omega^2}(
\tilde{R}+6\tilde{\square}\omega-6\tilde{g}^{\mu\nu}\partial_{\mu}\omega
\partial_{\nu}\omega)-\frac{U}{\Omega^4}\right]+\int\mathrm{d}^4x\,{\cal L}_{\rm mat}(
\Omega^{-2}\,\tilde{g}_{\mu\nu},\psi)\,.
\end{equation}
One can see that for $F>0$, 
the conformal factor which brings the action into Einstein frame is $\Omega^2=e^{2\omega}=F $.
Since the scale factor has a kinetic term in this frame, we promote it to a scalar field $\phi$, 
defined by
\begin{equation}
    \phi(x)\equiv \sqrt{6} \,\omega(x)\,M_P\,.
\end{equation}
Now we have all the necessary ingredients to write the action in the Einstein frame, where it takes the form \cite{DeFelice:2010aj}
\begin{equation}
\label{eq:EinstAction}
\tilde{S}=\int\mathrm{d}^{4}x\sqrt{-\tilde{g}}\left[\frac12\,M_P^2 \tilde{R}
+\frac{1}{2}\tilde{g}^{\mu\nu}\partial_{\mu}\phi\partial_{\nu}\phi-V(\phi)
\right]+\int\mathrm{d}^{4}x\,{\cal L}_{\rm mat}(F^{-1}(\phi)\tilde{g}_{\mu\nu},\psi)\,,
\end{equation}
with 
\begin{equation}
    V(\phi)=\frac{U}{F^{2}} \,.
\end{equation}
Note that the scalar field $\phi$ is canonically normalised.
For ease of notation let us name the scalar part of the Lagrangian as $\mathcal L_\phi$ so that $\mathcal{L}= \sqrt{-\tilde{g}} \left(M_P^2 \tilde{R}/2 + \mathcal{L}_\phi\right)+\mathcal{L}_{\rm mat}$.

\subsection{Inflationary plateau}

\begin{figure}
\centering
\includegraphics[scale=1]{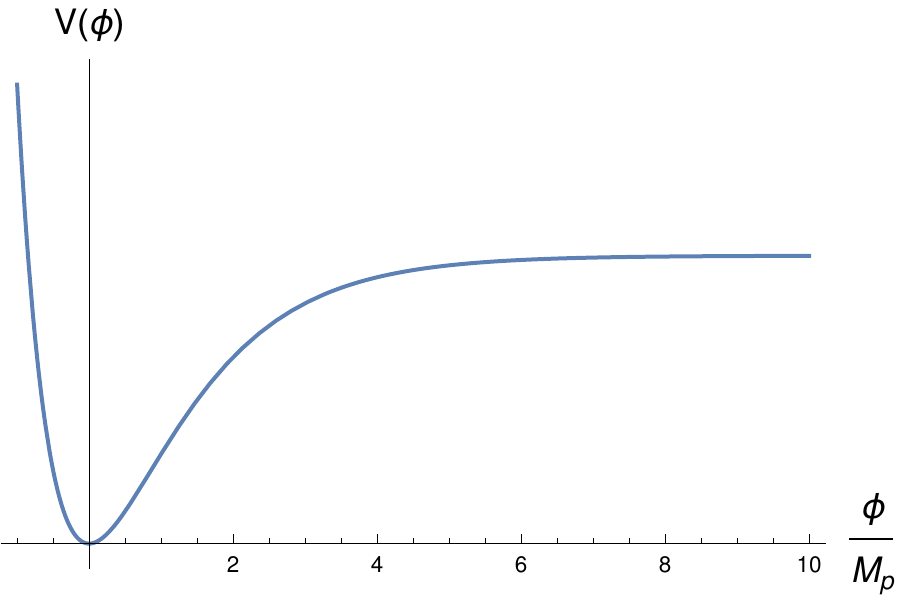}
\caption{Starobinsky potential \eqref{eq:StaroPot}. Inflation occurs for trans-Planckian values of the field $\phi$.}
\label{fig:StaroPot}
\end{figure}

To determine the scalar potential, note that the scalar field can also be expressed as 
\begin{equation}
\frac{\phi}{M_P}=\sqrt{\frac{3}{2}}  \ln F=
\sqrt{\frac{3}{2}}\,\ln\left(1+2\,\frac{R}{M^2}\right)\,.
\end{equation}
Then the scalar potential of the Starobinsky model gives the Starobinsky inflaton potential
\begin{equation}
\label{eq:StaroPot}
V(\phi)= \frac{R^2 M_P^2}{2 M^2}\left(1+2\,\frac{R}{M^2}\right)^{-2} =\frac18\,M^2 M_P^2 \left(1-e^{-\sqrt{2/3}\phi/M_P}\right)^{2}\,.
\end{equation}
This potential is plotted in Fig. \ref{fig:StaroPot}. From the form of the potential, we recognise two phases of evolution of the scalar field. Slow-roll inflation is realised for large values of the scalar field, $\phi\gg M_P$, where $V(\phi)\simeq M^2 M_P^2/8$. Here the potential is sufficiently flat to drive an epoch of accelerated expansion of the universe. More precisely, inflation ends around $\phi \simeq M_P$ and $50$-$60$ efoldings of inflation are realised around $\phi \simeq 5$-$6\,M_P$. Note that the WMAP normalisation of the CMB anisotropies constraints $M$ to be $M\simeq 10^{13}$ GeV. After this phase, the regime $\phi\ll M_P$ takes place, where $V(\phi)\simeq 3 M^2\phi^{2}$ and the field oscillates around $\phi=0$ leading to the reheating process.

\subsection{Coupling to matter}

Through the conformal transformation of the metric, the scalar field $\phi$ is directly coupled to matter in the Einstein frame \cite{DeFelice:2010aj}. We can see this better after obtaining an equation of motion for $\phi$. By taking the variation of \eqref{eq:EinstAction} with respect to $\phi$ we obtain 
\begin{equation}
    -\partial_{\mu}\left(\frac{\partial(\sqrt{-\tilde{g}}\,{\cal L}_{\phi})}{
\partial(\partial_{\mu}\phi)}\right)+\frac{\partial(\sqrt{-\tilde{g}}\,{\cal L}_
{\phi})}{\partial\phi}+\frac{\partial{\cal L}_{\rm mat}}{\partial\phi}=0\,,
\end{equation}
which simplifies to
\begin{equation}
\tilde{\square}\phi-\partial_\phi V+\frac{1}{\sqrt{-\tilde{g}}}\frac{\partial{\cal L}_{\rm mat}}{\partial\phi}=0\,.
\end{equation}
The matter term $\partial{\mathcal L}_{\rm mat}/\partial\phi$ is given by
\begin{equation}
    \frac{\partial{\cal L}_{\rm mat}}{\partial\phi}=\frac{\delta{\cal L}_{\rm mat}}{\delta g^{\mu\nu}}\frac{\partial g^{\mu\nu}}{\partial\phi}=\frac{1}{F(\phi)}\frac{\delta{\cal L}_{\rm mat}}{\delta\tilde{g}^{\mu\nu}}\frac{\partial(F(\phi)\tilde{g}^{\mu\nu
})}{\partial\phi}=-\sqrt{-\tilde{g}}\,\frac{\partial_\phi F}{2F}\,\tilde{T}_{\mu\nu}^{({\rm mat})}\tilde{g}^{\mu\nu}\,,
\end{equation}
where $\tilde{T}_{\mu\nu}^{({\rm mat})}$ is the energy-momentum tensor for matter 
\begin{equation}
\tilde{T}_{\mu\nu}^{({\rm mat})}\equiv \frac{2}{\sqrt{-\tilde{g}}}\frac{\delta{\cal L}_{\rm mat}}{\delta\tilde{g}^{\mu\nu}}\,.
\end{equation}
With this we obtain the field equation in the Einstein frame
\begin{equation}
\tilde{\square}\phi-\partial_\phi V + \frac{y_\phi}{M_P}\,\tilde{T}^{({\rm mat})} =0\,,
\end{equation}
with the Yukawa coupling between $\phi$ and matter given as 
\begin{equation}
y_\phi\equiv-\frac12\,M_P\, \partial_\phi \ln F=-\frac12\,M_P\frac{\partial_\phi (e^{\sqrt{\frac{2}{3}}\phi/M_P}) }{e^{\sqrt{\frac{2}{3}}\phi/M_P}}=-\frac{1}{\sqrt{6}}\,.
\end{equation}

This shows that the scalar field $\phi$ is directly coupled to matter with a universal coupling constant $y_\phi=-1/\sqrt{6}$. Alternatively, the coupling can be read off directly from the action 
\begin{equation}
\mathcal{L}_{\rm mat} \supset   -y_\phi \,\frac{m_\psi}{M_P}\, \phi \bar{\psi}\psi\,, 
\end{equation}
which gives the same universal result. This is also the way we shall compute the coupling for stringy models in Sec. \ref{sec:stringy}.

\subsection{General $f(R)$ expansion}
\label{Genf(R)}

\begin{figure}
\centering
\includegraphics[scale=1]{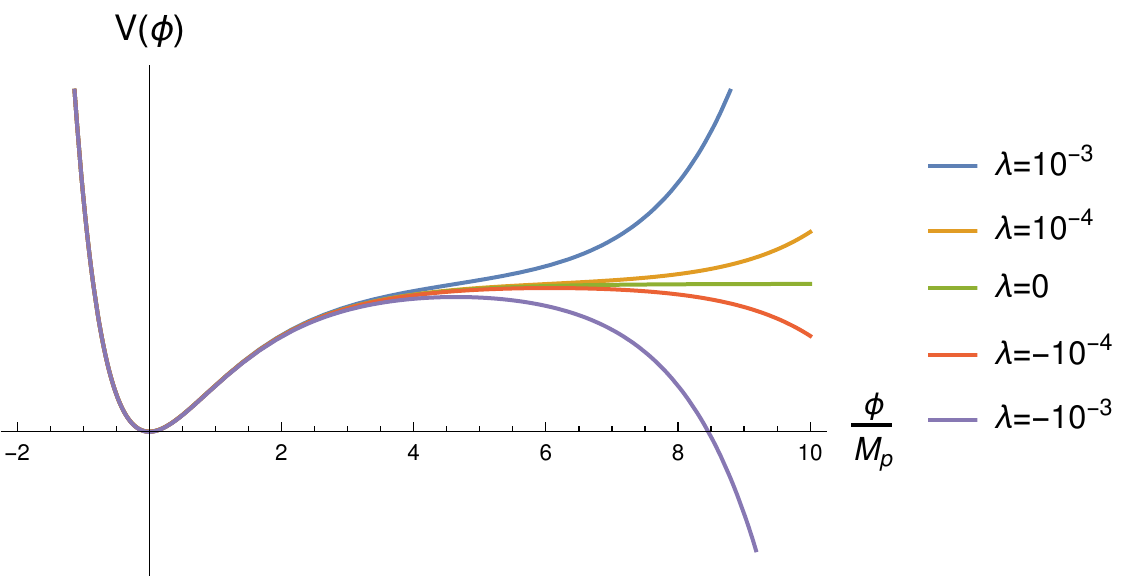}
\caption{Starobinsky potential with $R^4$ corrections given by \eqref{CorrPot}.}
\label{fig:StaroPotCorr}
\end{figure}

Let us now consider a more general $f(R)$ theory where the function $f(R)$ is analytic around zero, and so can be expanded in a series in $R^2/M$. This is motivated by the fact that if Starobinsky inflation arises from a more fundamental UV complete theory, we expect it to be an effective field theory obtained after integrating out massive modes below some high scale $M$ (which can still be several orders of magnitude below the Planck scale). In this effective theory we expect all possible terms compatible with the underlying symmetries and particle content. Hence we expect not just terms involving the Ricci scalar, but also contributions which depend on the Ricci and the Riemann tensors. Setting this issue aside, and focusing just on an effective $f(R)$ theory, we generalise (\ref{eq:f(R)}) by a series expansion in $R^2/M$ of the form (see \cite{Huang:2013hsb, Cheong:2020rao, Ivanov:2021chn, Lee:2023wdm}
 for similar studies):
\begin{equation}
f(R) = R+ \frac{R^2}{M^2} + \mu\, \frac{R^3}{M^4} + \lambda\,\frac{R^4}{M^6} + ...  
\label{fGen}
\end{equation}
The resulting scalar potential in Einstein frame would look like
\begin{equation}
V = \frac18 M^2 M_P^2\left[\left(1-e^{-\sqrt{\frac23}\frac{\phi}{M_P}}\right)^2 + \lambda\,e^{\sqrt{\frac23}\frac{\phi}{M_P}} + ...\right]   
\label{CorrPot}
\end{equation}
where we have set $\mu=0$ since in the following we shall focus on supersymmetric models where this coefficient vanishes. A working Starobinsky inflation model requires $\lambda \ll 1$ or, in other words, that the $R^4$ term is suppressed by an effective scale $\tilde{M} = M \,\lambda^{-6} \gg M$ which is well above $M$. Fig. \ref{fig:StaroPotCorr} shows the effect of $R^4$ corrections for different values of $\lambda$. It is easy to infer that these corrections do not ruin $50$-$60$ efoldings of standard Starobinsky inflation only if $|\lambda|<10^{-4}$.

This has to be the case, more in general, for all higher order terms $R^n$ with $n\geq 5$. Thus, a working Starobinsky inflation model requires the presence of at least two different suppression scales, which might be hard to justify from a low-energy perspective. The idea is that $\tilde{M}$ could correspond to the Planck scale so that (\ref{fGen}) behaves as
\begin{equation}
f(R) = R+ \frac{R^2}{M^2} \left(1+ c_1\frac{R}{M_P^2} + c_2\frac{R^2}{M_P^4} + ... \right) 
\label{fGenNew}
\end{equation}
where the $c_i$ are now $\mathcal{O}(1)$ coefficients. While this expression can certainly be postulated from an effective field theory point of view, it definitely deserves an explanation from a more fundamental theory which is valid at higher energy scales.

\section{A stringy embedding?}
\label{sec:stringy}

In this section we will explore the possibility to embed Starobinsky inflation in string theory by either deriving higher curvature corrections from ten dimensions in Jordan frame, or by reproducing the correct potential and Yukawa coupling of the Starobinsky scalar in Einstein frame. 

\subsection{Higher curvatures from ten dimensions}

Let us investigate if the structure outlined in (\ref{fGenNew}) with two suppression scales can be reproduced from string theory following \cite{Burgess:2016owb}. The low-energy ten-dimensional action below the string scale takes the schematic form
\begin{equation}
\frac{1}{\alpha'^4} \int d^{10}\xi \sqrt{-G}\, e^{-2\Phi} \left[ \mathcal{R} + k_1\, \alpha'\mathcal{R}^2 + k_3\,\alpha'^3 \mathcal{R}^4 + .... \right]
\end{equation}
where $k_1\neq 0$ for heterotic and type I strings, while $k_1=0$ for type IIB string theory, and $k_3\sim\mathcal{O}(1)$ for all string theories. Let us set the dilaton $\Phi$ and the internal volume $\mathcal{V}$ to their vacuum expectation values, given respectively by 
\begin{equation}
e^{\Phi_0} = g_s\qquad\text{and}\qquad \int d^6y \,\sqrt{-g_{(6D)}} = \mathcal{V}\,\alpha'^3\,.
\end{equation}
Dimensional reduction to four dimensions then gives (showing only the terms which involve four-dimensional curvatures)
\begin{equation}
\frac{M_P^2}{2}\, \int d^4 x \sqrt{-g} \left[ R + \frac{R^2}{M_s^2} \left(k_1 + \frac{\tilde{k}_1}{\mathcal{V}^{2/3}}+ ... \right) + k_3\,\frac{R^4}{M_s^6} + .... \right]
\end{equation}
where $\tilde{k}_1\sim\mathcal{O}(1)$ and
\begin{equation}
M_s \simeq \frac{1}{\sqrt{\alpha'}}\,,\qquad\text{and}\qquad M_P^2 \simeq M_s^2 \,\frac{\mathcal{V}}{g_s^2}\,.
\label{MassRelations}
\end{equation}
This analysis shows that Starobinsky inflation is very unlikely to emerge in string theory from the dimensional reduction of ten-dimesional higher derivative terms for the following reasons:
\begin{itemize}
\item \emph{Heterotic string:} in this case $k_1\neq 0$ and the Starobinsky suppression scale $M$ is identified with the string scale $M_s$, $M\simeq M_s$. Due to the relation (\ref{MassRelations}), it might not seem difficult to achieve $M_s\simeq 10^{13}$ GeV which is necessary to match the observed amplitude of the density perturbations if $g_s \ll 1$ and $\mathcal{V}\gg 1$, which are the limits where the supergravity approximation is under control. However the string scale is tied to the gauge coupling of the visible sector (generically a GUT theory) since $\alpha^{-1}_{\rm GUT} \simeq \mathcal{V}/g_s^2$. To avoid hyperweak couplings, the string scale is therefore around the GUT scale, $M_s\simeq 10^{16}$ GeV, which is too high to reproduce $M_s\simeq 10^{13}$ GeV. Let us stress that, even ignoring this issue, the $R^4$ term would still be suppressed by the same scale, the string scale, resulting in a positive exponential that spoils the inflationary plateau.

\item \emph{Type IIB string:} In this case $k_1=0$, and so the Starobinsky suppression scale $M$ has to be identified with
\begin{equation}
M\simeq M_s \mathcal{V}^{1/3} =  \frac{g_s M_P}{\mathcal{V}^{1/6}}\,,
\end{equation}
which could again reproduce $M\simeq 10^{13}$ GeV for $g_s \ll 1$ and $\mathcal{V}\gg 1$. In type IIB the overall internal volume is decoupled from the visible sector gauge coupling since the Standard Model (or genelisations thereof) is realised on stacks of D3 or D7-branes which are localised in the extra-dimensions. However the mass scale which suppresses the $R^4$ terms would be given simply by the string scale, and so it would be even smaller than the one suppressing the $R^2$ term. This implies that the positive exponential arising from the $R^4$ term would destroy inflation very quickly.
\end{itemize}

\subsection{Moduli inflation and Yukawa couplings}

String theory proposes various candidate scalar fields that could drive inflation. Many interesting scenarios have been explored in the past years (see \cite{Cicoli:2023opf}). Inflationary solutions have been developed by employing open string fields controlling the position of either warped, unwarped or relativistic branes. Alternatively axions have also been proposed as candidates to drive inflation thanks to their perturbative shift symmetry. Most importantly for us, the possibility that K\"ahler moduli, naturally arising in string compactifications, could be the inflaton field has also been considered in a vast literature \cite{Conlon:2005jm,Bond:2006nc, Blanco-Pillado:2009dmu, Cicoli:2008gp, Cicoli:2016chb, Broy:2015zba, Cicoli:2011ct, Cicoli:2015wja, Conlon:2008cj, Antoniadis:2018ngr, Antoniadis:2020stf, Cicoli:2016xae, Cicoli:2017axo, Cicoli:2020bao, Bhattacharya:2020gnk, Broy:2014xwa} since these fields (particularly the modes orthogonal to the overall volume) enjoy an effective and approximate non-compact shift symmetry \cite{Burgess:2014tja, Burgess:2016owb}. In what follows we will focus on K\"ahler moduli since in supergravity effective theories they naturally admit an exponential potential that is the typical feature of Starobinsky inflation. 

Our goal here is to explore the possibility that any of these K\"ahler moduli can resemble the Starobinsky field and therefore be eligible as a candidate for driving Starobinsky inflation. The first step is to verify that the coupling of this candidate modulus to matter coincides with the coupling to matter of the Starobinsky field, which is equal to $y_\phi=-1/\sqrt{6}$. This will be our primary objective in the following discussion, where we will first consider as possible candidates the volume modulus and blow-up modes. Subsequently, we analyse the form of the scalar potential to check if it resembles the Starobinsky potential. We will also examine a particular class of constructions known as fibre inflation. Again, we will focus on the computation of the couplings to matter and of the scalar potential of bulk moduli appearing in fibred Calabi-Yau models of string inflation.

\subsubsection{Volume modulus and blow-up modes}
\label{subsec:vol}

Due to the presence of the compactification volume in all the four-dimensional effective theories of string compactifications, it is natural to ask if it could be the field driving slow-roll inflation. The most concrete models have been constructed in the Large Volume Scenario (LVS) \cite{Balasubramanian:2005zx}. Following \cite{Conlon:2007gk}, we will consider compactifications on $\mathbb{P}^4_{[1,1,1,6,9]}$. This is the single-hole Swiss cheese geometry with two K\"ahler moduli, $T_b=\tau_b + i \theta_b$ and $T_s=\tau_s + i \theta_s$, and Calabi-Yau volume given by (up to an irrelevant overall $1/(9\sqrt{2})$ factor)
\begin{equation}
\label{eq:Volume} 
\mathcal{V}= \tau_b^{3/2} -\tau_s^{3/2} \,.
\end{equation}
The big modulus $\tau_b$ controls the size of the overall volume, while $\tau_s$ controls the size of the blow-up of a point-like singularity. The K\"ahler potential and the superpotential are given by (setting from now on $M_P=1$)
\begin{eqnarray}
K&=&-2\ln\left[\left(\tau_{b}^{3/2}-\tau_{s}^{3/2}\right)+\frac{\hat\xi}{2}\right] \\
W&=&W_{0}+A_{s}\,e^{-a_{s}T_{s}},
\end{eqnarray}
where $\hat\xi=\xi/g_s^{3/2}$, with $\xi=\zeta(3)\chi(M)/(2\pi)^{3}$ a constant controlling the size of $\mathcal{O}(\alpha'^3)$ corrections \cite{Becker:2002nn}. Here $\zeta(3)\simeq 1.20$ is Ap\'ery's constant and $\chi(M)$ is the Euler number of the Calabi-Yau manifold.

We will assume that the procedure of moduli stabilisation has been performed, giving the moduli a potential and a mass. Expanding the fields in $\tau_i=\langle \tau_i \rangle + \delta \tau_i$, we can write, around the minimum of the moduli potential, the following Lagrangian \cite{Conlon:2007gk} 
\begin{equation}
\mathcal{L}=K_{i\bar{j}}\partial_{\mu}(\delta\tau_{i})\partial^{\mu}
(\delta\tau_{j})-V_0-(\mathcal{M}^{2})_{ij}(\delta\tau_{i})(\delta\tau_{j})-\mathcal{O}(\delta\tau^{3})- \frac{1}{4 g_\SM^2} F_{\mu\nu}F^{\mu\nu}\,,
\end{equation}
where the Standard Model gauge coupling $g_\SM$ can take three different expressions depending on the brane realisation of the visible sector: ($i$) $g^{-2}_\SM = \tau_s$ for D7-branes wrapped around the blow-up mode $\tau_s$; ($ii$) $g^{-2}_\SM = \tau_b$ for D7-branes wrapped around the big four-cycle $\tau_b$ (in this case the overall volume cannot be too large in order to avoid a hyperweak Standard Model gauge coupling); ($iii$) $g^{-2}_\SM = s$ for D3-branes at singularities (where $s$ is the dilaton with $\langle s \rangle = g_s^{-1}$).

It is useful to rewrite the Lagrangian in terms of mass eigenstates $\sigma$ and $\phi$ (the heavy and the light field, respectively), defined by
\begin{equation}
    \begin{pmatrix}\delta\tau_{b}  \\
\delta\tau_{s}\end{pmatrix}=\Bigg{(}v_{\sigma}\Bigg{)}\frac{\sigma}{\sqrt{2}}
+\Bigg{(}v_{\phi}\Bigg{)}\frac{\phi}{\sqrt{2}}\;,
\end{equation}
where $v_{\sigma}$ and $v_{\phi}$ are the normalised eigenvectors of $(K^{-1})_{i\bar{j}}(\mathcal{M}^2)_{\bar{j}k}$. 
In the following, we will need an expression for the original fields $\delta \tau_b$ and $\delta \tau_s$ in terms of $\sigma$ and $\phi$. These relations are given at leading order by \cite{Conlon:2007gk}
\begin{eqnarray}
\delta\tau_{b}&=&\displaystyle\left(\sqrt{6}\langle\tau_{b}\rangle^{1/4}\langle\tau_{s}\rangle^{3/4}
\right)\frac{\sigma}{\sqrt{2}}+
\left(\sqrt{\frac{4}{3}}\langle\tau_{b}\rangle\right)\frac{\phi}{\sqrt{2}}\,
\sim\,\mathcal{O}\left(\mathcal{V}_0^{1/6}\right)\sigma\,+\,
\mathcal{O}\left(\mathcal{V}_0^{2/3}\right)\phi
\label{eq:BigSmall} \\
\delta \tau_s &=& \left(\frac{2\sqrt{6}}{3}\langle\tau_{b}\rangle^{3/4}\langle\tau_{s}\rangle^{1/4}\right)
\frac{\sigma}{\sqrt{2}}+\left(\frac{\sqrt{3}}{a_{s}}
\,\right)\frac{\phi}{\sqrt{2}}\,\sim\,\mathcal{O}\left
(\mathcal{V}_0^{1/2}\right)\sigma\,+\,\mathcal{O}\left(1\right)\phi\,,
\end{eqnarray}
with $\mathcal{V}_{0}=\langle\mathcal{V}\rangle\simeq \langle \tau_b\rangle^{3/2} \gg 1$. 

\subsubsection*{Couplings to matter}
\label{subsubsec:vol_coupling}

Let us now compute the couplings between moduli and chiral matter fields. We consider the following relevant terms of the supergravity Lagrangian \cite{Conlon:2007gk}
\begin{equation}
\label{eq:SugraLagr}
    \mathcal{L}=K_{\bar{\psi}\psi} i \bar{\psi}\gamma^{\mu}\partial_{\mu}\psi+K_{H\bar{H}}
\partial_{\mu}H\partial^{\mu}\bar{H}-e^{K/2} y_\H H\bar{\psi}\psi\;.
\end{equation}
The K\"ahler metric for the chiral matter fields is given by 
\begin{equation}
K_{\bar{\psi}\psi}\sim K_{\bar{H}H}\sim\frac{\tau_{s}^{\lambda_s}}{\tau_{b}^{\lambda_b}}=K_{0}\left(1+
\lambda_s\frac{\delta\tau_{s}}{\langle\tau_{s}\rangle}-\lambda_b\frac{\delta\tau_{b}}
{\langle\tau_{b}\rangle}+\ldots\right)\;,
\end{equation}
with $K_0\equiv\langle\tau_{s}\rangle^{\lambda_s}/\langle\tau_{b}\rangle^{\lambda_b}$ and, depending on the detailed origin of the matter fields as open strings on branes, one can have \cite{Aparicio:2008wh} 
\begin{alignat}{3}
&\lambda_s = 1,1/2,0\,,\qquad &&\lambda_b = 1 &&\qquad\text{if}    \quad g^{-2}_\SM = \tau_s\,, 
\label{SMonS} \\
&\lambda_s = 0\,,\qquad &&\lambda_b = 1,1/2,0 &&\qquad\text{if}    \quad g^{-2}_\SM = \tau_b\,,
\label{SMonB} \\
&\lambda_s = 0\,,\qquad &&\lambda_b = 1 &&\qquad\text{if}    \quad g^{-2}_\SM = s\,. 
\label{SMonD3}
\end{alignat}
More precisely, when the Standard Model is realised with D7-branes wrapping the four-cycle $\tau_i$, the modular weight $\lambda_i=1$ corresponds to open strings living within the D7-worldvolume, $\lambda_i=1/2$ to matter fields at the intersection between two stacks of D7-branes, and $\lambda_i=0$ to open string moduli controlling the position of D7-branes.

Moreover we will use the expansion \cite{Conlon:2007gk}
\begin{equation}
e^{K/2}=\frac{1}{\mathcal{V}}\simeq\frac{1}{\tau_{b}^{3/2}-\tau_s^{3/2}}=\frac{1}{\mathcal{V}_{0}}\left(1-\frac{3}{2}\left(\frac{\delta\tau_{b}}{\langle\tau_{b}\rangle}\right)+\ldots\right)\,.
\end{equation}
In terms of these expressions, we can now write the Lagrangian \eqref{eq:SugraLagr} as
\begin{eqnarray}
\mathcal{L}&=&K_{0}\,i \bar{\psi}\gamma^{\mu}\partial_{\mu}\psi+ K_{0}\,\partial_{\mu}H
\partial^{\mu}\bar{H}-\frac{y_\H}{\mathcal{V}_{0}}  H\bar{\psi}\psi+\left(\lambda_s\left(\frac{\delta\tau_{s}}{\langle\tau_{s}\rangle}\right)-\lambda_b\left(\frac{\delta\tau_{b}}{\langle\tau_{b}\rangle}\right)\right)K_{0}\,i\bar{\psi}\gamma^{\mu}\partial_{\mu}\psi \nonumber \\[3pt]
&&+\left(\lambda_s\left(\frac{\delta\tau_{s}}{\langle\tau_{s}
\rangle}\right)-\lambda_b\left(\frac{\delta\tau_{b}}{\langle\tau_{b}\rangle}\right)
\right)K_{0}\,\partial_{\mu}H\partial^{\mu}\bar{H}+\frac{3}{2}\left(\frac{\delta\tau_{b}}{\langle\tau_{b}\rangle}\right)\frac{y_\H}{\mathcal{V}_{0}} H
\bar{\psi}\psi\,.
\end{eqnarray}
The next step is to canonically normalise the fields and impose electroweak symmetry breaking. The Higgs will acquire a vacuum expectation value and the following fermion mass will be generated
\begin{equation}
     y_\H \langle H_{\textrm{c}} \rangle = m_\psi\;.
\end{equation}
Therefore the Lagrangian describing the cubic couplings between matter fields and moduli looks like
\begin{equation}
\label{eq:EffLagr}
\mathcal{L}_{\rm cubic} = \left(\lambda_s\frac{\delta
\tau_{s}}{\langle\tau_{s}\rangle}-\lambda_b\frac{\delta\tau_{b}}{\langle\tau_{b}\rangle
}\right)\bar{\psi}\,(i\gamma^{\mu}\partial_{\mu}-m_{\psi})\,\psi+\left[\lambda_s\frac{
\delta\tau_{s}}{\langle\tau_{s}\rangle}+\left(\frac32-\lambda_b\right)\frac{\delta\tau_{b}}{\langle\tau_{b}\rangle}\right]m_{\psi}\,\bar{\psi}\psi\,,
\end{equation}
where only the last term gives a non-zero contribution after the equations of motion are taken into account. Our first goal is to prove that the volume modulus $\tau_b$ has the right coupling to matter. 

Let us first remark that from \eqref{eq:BigSmall}, it is possible to notice that $\tau_b$ is mostly given by $\phi$ while $\tau_s$ is mostly $\sigma$. To get the coupling, let us now consider the last term of \eqref{eq:EffLagr}, in particular focussing on the $\tau_b$ part
\begin{equation}
\label{eq:BigCoupling}
\mathcal{L}_{\phi \bar{\psi} \psi} \simeq \left(\frac32-\lambda_b\right)  \frac{\delta \tau_b}{\langle \tau_b \rangle}\, m_\psi \bar{\psi} \psi\;. 
\end{equation}
Given that the light field $\phi$ is dominantly the volume modulus, we can reduce \eqref{eq:BigSmall} to 
\begin{equation}
\frac{\delta \tau_b}{\langle \tau_b \rangle} \simeq \sqrt{\frac{2}{3}}\,\phi\,.
\end{equation}
Using this, \eqref{eq:BigCoupling} can be written as
\begin{equation}
\mathcal{L}_{\phi \bar{\psi} \psi} \simeq -\sqrt{\frac23}\left(\lambda_b-\frac32\right) \,m_\psi\, \phi \bar{\psi} \psi\,,
\end{equation}
from which we can easily read off the coupling between the light field $\phi$ and matter, which is equal to 
\begin{equation}
y_\phi=\sqrt{\frac23}\left(\lambda_b-\frac32\right) = -\frac{1}{\sqrt{6}}\qquad\text{if}\quad \lambda_b=1\,.
\label{yphi}
\end{equation}
As shown above, (\ref{yphi}) reproduces the correct conformal coupling to matter of Starobinsky inflation, $y_\phi = -1/\sqrt{6}$, if $\lambda_b=1$. As can be seen from (\ref{SMonS}), (\ref{SMonB}) and (\ref{SMonD3}), this is indeed very often the case since $\lambda_b=1$ if the Standard Model lives on D7-branes wrapped around the blow-up mode, on D3-branes but also on D7-branes wrapped around the big four-cycle $\tau_b$ provided the matter fields originate from the
reduction of gauge fields within the D7-worldvolume \cite{Aparicio:2008wh}. In our limit, the light field $\phi$ is basically the volume modulus $\tau_b$. The fact that the volume modulus has the same coupling to matter as the Starobinsky field represents a first step towards a possible identification between the two. 

Let us now present the coupling to matter of the heavy field $\sigma$, showing that it is different from $-1/\sqrt{6}$, therefore excluding the possibility for the $\tau_s$ modulus to be the scalar field responsible for Starobinksy inflation. To proceed, let us consider the $\tau_s$ part in the last term of \eqref{eq:EffLagr}
\begin{equation}
\mathcal{L}_{\sigma \bar{\psi} \psi} \simeq \lambda_s \left( \frac{\delta \tau_s}{\langle \tau_s \rangle} \right) m_\psi \, \bar{\psi} \psi\;.
\end{equation}
From the expansion \eqref{eq:BigSmall}, we can see that it is possible to make the following approximation
\begin{equation}
\frac{\delta \tau_s}{\langle \tau_s \rangle} \simeq \sqrt{\mathcal{V}_0}\,\sigma\;.
\end{equation}
We conclude that
\begin{equation}
\mathcal{L}_{\sigma \bar{\psi} \psi} \simeq \lambda_s  \sqrt{\mathcal{V}_0} \,m_\psi\, \sigma \bar{\psi} \psi\,.
\end{equation}
The general classification (\ref{SMonS}), (\ref{SMonB}) and (\ref{SMonD3}) shows that $\lambda_s$ is non-zero only when the Standard Model lives on D7-branes wrapped around $\tau_s$. However in this case the absolute value of the Yukawa coupling to matter of the heavy field $\sigma$ (mainly the small modulus $\tau_s$) is much larger than $1/\sqrt{6}$ since $\mathcal{V}_0\gg 1$. Note also that $\sigma$ is approximately decoupled from ordinary matter when the Standard Model is realised via D7-branes on $\tau_b$ or via D3-branes at singularities since these cases are characterised by $\lambda_s=0$. A non-zero Yukawa coupling would be generated by considering the term in \eqref{eq:EffLagr} proportional to $\delta \tau_b$ and then exploiting the dependence of $\delta \tau_b$ on $\sigma$ in (\ref{eq:BigSmall}). However it is easy to check that this would induce a very suppressed coupling proportional to $\mathcal{V}_0^{-1/2}\ll 1$. We therefore conclude that $\sigma$ cannot be identified with the Starobinksy field.

\subsubsection*{Scalar potential}
\label{subsubsec:vol_scalar_pot}

Although the K\"ahler modulus $\tau_b$ controlling the overall volume of the Calabi-Yau manifold has the right coupling to matter, it is well known that its scalar potential is not suitable for a slow-roll inflationary behaviour. Recall that the Starobinsky potential has the form (\ref{eq:StaroPot}) with a slow-roll plateau at large values of $\phi$.

Such a constant behaviour of the inflationary potential can never be reproduced for the volume mode since any contribution to the scalar potential has to depend on $\mathcal{V}$ due to the Weyl rescaling to express the four-dimensional action in Einstein frame. Alternatively, this can also be seen from the general expression of the supergravity scalar potential that is proportional to $e^K = \mathcal{V}^{-2}$. This has to be the case since the potential has to go to zero in the infinity volume limit, $\mathcal{V}\to \infty$, in order to recover a ten-dimensional Minkowski solution in the decompactification regime. 

Due to these very general considerations, at large field values, the potential of the volume modulus will feature a runaway behaviour, instead of a Starobinsky-like plateau, regardless of the effects (perturbative or non-perturbative) which break the no-scale structure and generate the potential for $\mathcal{V}$. Without loss of generality, we therefore illustrate this claim by focusing on the simplest LVS scenario where the (uplifted) scalar potential reads
\begin{equation}
\label{eq:SugraScalarPot}
    V=\frac{8\sqrt{\tau_{s}}(a_{s}A_{s})^{2}e^{-2a_{s}\tau_{s}}}{3\mathcal{V}}-
\frac{4W_{0}a_{s}A_{s}\tau_{s}e^{-a_{s}\tau_{s}}}{\mathcal{V}^{2}}+\frac{3\hat\xi W
_{0}^{2}}{4\mathcal{V}^{3}} +V_{\rm up}\;,
\end{equation}
where, again without loss of generality, we parameterised the uplift potential as $V_{\rm up} = {\delta}/{\mathcal V^\alpha}$ with $\alpha<3$. A single-field potential for the volume modulus can be obtained upon integrating out $\tau_s$
\begin{equation}
V = \frac{\hat\xi W_{0}^{2}}{2\mathcal{V}^{3}} \left(1-\frac{8\,(\textrm{ln}\,\mathcal{V})^{3/2} }{3\,\hat\xi a_s^{3/2}} \right)
+ \frac{\delta}{\mathcal V^\alpha}\;.
\label{eq:BigModulusPotential}
\end{equation}
This expression depends only on the volume $\mathcal{V}$, which is related to the modulus $\tau_b$ and to the canonically normalised field $\phi$ through
\begin{equation}
\mathcal{V}^{2/3}\simeq \tau_b = e^{\sqrt{\frac{2}{3}}\phi}\;.
\end{equation}
In terms of the canonical field $\phi$, the volume mode potential reads
\begin{equation}
V(\phi)=\frac{\hat\xi W_{0}^{2}}{2}\left(1-\frac{8\left( \sqrt{\frac{3}{2}}\phi\right)^{3/2}}{3\,\hat\xi a_s^{3/2}} \right)\,e^{-\sqrt{\frac{27}{2}}\phi}     + \delta \,e^{-\alpha\sqrt{\frac32} \phi} \;.
\label{eq:canonicalVolModPotential}
\end{equation}

\begin{figure}[t]
\centering
\includegraphics[scale=1]{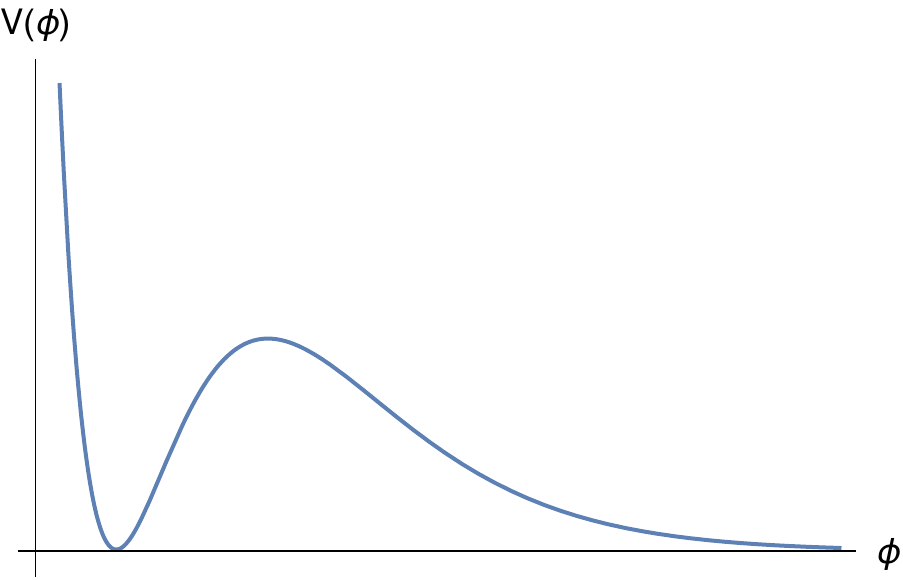}
\caption{The volume modulus potential \eqref{eq:canonicalVolModPotential} for a parameter choice that realises an uplift to Minkowski, featuring the typical runaway towards large volume.}
\label{fig:PotentialUplift}
\end{figure}

This potential is shown in Fig. \ref{fig:PotentialUplift}. As anticipated, due to the volume dependence of the uplift term, the potential does not have the necessary plateau of the Starobinsky potential, featuring instead a typical runaway at large field values. Thus the volume modulus $\tau_b$ cannot drive Starobinksy inflation. 

On the upside, since the uplift depends on the volume only, the uplifting contribution is constant in other directions of moduli space. If we can find another modulus with the correct coupling to the matter sector, this could provide the plateau we need.

\subsubsection{Fibre moduli}
\label{subsec:fibre}

In the context of LVS moduli stabilisation in type IIB Calabi-Yau orientifold compactifications, another particular realisation of inflation is given by the class of constructions known as fibre inflation \cite{Cicoli:2008gp,Burgess:2016owb,Cicoli:2016xae,Cicoli:2017axo,Cicoli:2020bao,Bhattacharya:2020gnk,Cicoli:2016chb,Broy:2015zba}. We now briefly discuss its setup following \cite{Cicoli:2008gp,Burgess:2016owb,Cicoli:2018cgu}.

The main idea behind fibre inflation is to use as the inflaton one of the K\"ahler moduli whose potential is generated by perturbative (in $\alpha'$ or $g_s$) corrections to the K\"ahler potential \cite{Berg:2005ja, vonGersdorff:2005bf, Berg:2007wt, Cicoli:2007xp, Ciupke:2015msa, Cicoli:2021rub} which are subdominant with respect to the leading $\mathcal{O}(\alpha'^3)$ term \cite{Becker:2002nn} that we have considered to stabilise the volume modulus. The simplest construction uses a Calabi-Yau manifold which is a K3 fibration over a $\mathbb{P}^1$ base. For concreteness, let us consider the $\mathbb{P}^4_{[1,1,2,2,6]}$ geometry. The model has two K\"ahler moduli $\tau_1$ and $\tau_2$. We require also the existence of a third K\"ahler modulus which is implemented by the presence of an additional blow-up cycle, whose volume we denote by $\tau_s$ \cite{Cicoli:2011it, Cicoli:2018tcq}. To summarise, this model (and also fibre inflation models in general) involves a Calabi-Yau manifold with at least three K\"ahler moduli (denoting them by $T_i=\tau_i + i\theta_i$, with $\tau_i$ the geometric modulus and $\theta_i$ its axionic partner):
\begin{itemize}
\item $T_1=\tau_1 + i \theta_1$. The geometric modulus $\tau_1$ corresponds to the volume of a K3 fibre over a $\mathbb{P}^1$ base.

\item $T_2=\tau_2 + i \theta_2$, where $\tau_2$ is the volume of a divisor which contains the $\mathbb{P}^1$ base.

\item $T_s=\tau_s + i \theta_s$. As in the Swiss cheese example, $\tau_s$ corresponds to the volume of a blow-up cycle (the `hole' of the Swiss cheese).
\end{itemize}
There are some differences with respect to the previously considered LVS model. Here, $\tau_1$ is stabilised at subleading order due to string loop or higher $\alpha'$ corrections to the K\"ahler potential. The overall volume $\mathcal{V}$ is mainly controlled by $\tau_2$ which is stabilised at leading order by $\mathcal{O}(\alpha'^3)$ corrections. The most important difference is that the volume modulus $\mathcal{V}$ acts only as a spectator during inflation and it is heavier than the inflaton. For what concerns $\tau_s$, it is fixed by non-perturbative corrections to the superpotential $W$ and it is heavier than the inflaton $\tau_1$ during inflation.

In terms of these moduli, the compactification volume is given by
\begin{equation}
\mathcal{V}= \kappa_0\left(\sqrt{\tau_1}\tau_2 - \kappa_s \tau_s^{3/2}\right) \,,
\end{equation}
with $\kappa_0$ and $\kappa_s$ two model-dependent constants of order one, which are determined by the Calabi-Yau intersection numbers. Without loss of generality, we shall set $\kappa_0=\kappa_s=1$.

\subsubsection*{Scalar potential}

The idea behind fibre inflation is to rely on perturbative corrections to the effective action. Before including these, it may be useful to summarise the procedure for obtaining the leading order scalar potential.

We compute the scalar potential by considering the leading $\alpha'$ corrections to the K\"ahler potential $K$, and non-perturbative corrections to the superpotential $W$
\begin{equation}
K=K_{0}+\delta K_{\alpha'}=-2\ln\left(\mathcal{V}+\frac{\hat\xi}{2}\right)\qquad\hbox{and}\qquad W=W_{0}+\sum_i A_i\,e^{-a_i T_i}\,.
\end{equation}
Since we are interested in the large volume regime where
\begin{equation}
\sqrt{\tau_1} \tau_2 \gg  \tau_s^{3/2}\,,
\end{equation}
we can actually neglect non-perturbative effects involving $\tau_1$ and $\tau_2$ (i.e. the $T_1$ and $T_2$ dependence in $W$) and keep only the $T_s$ dependence:
\begin{equation}
    W\simeq W_{0}+A_{s}\,e^{-a_{s}T_{s}}\,.
\end{equation}

Including an uplifting sector, the leading order scalar potential takes the form (\ref{eq:SugraScalarPot}). The key point of this construction is that this potential depends only on $\tau_s$ and $\mathcal{V}$. These get stabilised at (in the $a_s \tau_s \gg 1$ limit)
\begin{equation}
\langle\tau_{s}\rangle=\left(\frac{\hat\xi}{2}\right)^{2/3}
\qquad\hbox{and}\qquad\langle\mathcal{V}\rangle=\left(\frac{3}{4
a_{s}A_{s}}\right)W_{0}\,\sqrt{\langle\tau_{s}\rangle}\;e^{a_{s}\langle\tau_{s
}\rangle}\text{\ .}
\end{equation}
Most importantly, we have a completely flat direction in the $(\tau_1,\tau_2)$-plane, along which $\mathcal{V}$ is constant. 
This flat direction is lifted by subdominant contributions to the K\"ahler potential. In fibre inflation these are the corrections providing the leading terms in the inflaton potential.

Focusing on the version of fibre inflation where the inflaton potential arises from string loop effects \cite{Berg:2005ja, vonGersdorff:2005bf, Berg:2007wt, Cicoli:2007xp}, we have\footnote{Recall that string loops are perturbative corrections, and so we work in the regime $W_0 \gtrsim \mathcal{O}(1)$, where they become relevant.}
\begin{equation}
\delta V_{g_s}=\frac{W_0^2}{\mathcal{V}^2} \left( A\, \frac{g_s^2}{\tau_1^2} - \frac{B}{\sqrt{\tau_1} \mathcal{V}} + C\, \frac{g_s^2\tau_1}{\mathcal{V}^2}\right),
\end{equation}
where $A$, $B$ and $C$ are functions of the complex structure moduli which are expected to become $\mathcal{O}(1)$ constants after these moduli are frozen by background fluxes. Taking this contribution into account, the fibre modulus $\tau_1$ gets fixed at
\begin{equation}
    \langle \tau_1 \rangle \simeq g_s^{4/3} \langle \mathcal{V} \rangle^{2/3}\;.
\end{equation}

So, how can we achieve inflation? We have established the existence of an LVS minimum of the potential. The idea is to displace a field away from its minimum and explore the possibility of having an inflationary dynamics. Due to the fact that the potential for $\tau_1$ is flat if we do not include string loop corrections, this is the field we displace to drive inflation. Displacing $\tau_1$ from its minimum is equivalent to increasing the size of the K3 fibre while shrinking the base. If we consider $\tau_s$ and $\mathcal{V}$ stabilised at their minima, we can integrate them out to obtain a single-field potential.

\begin{figure}
\centering
\includegraphics[scale=1]{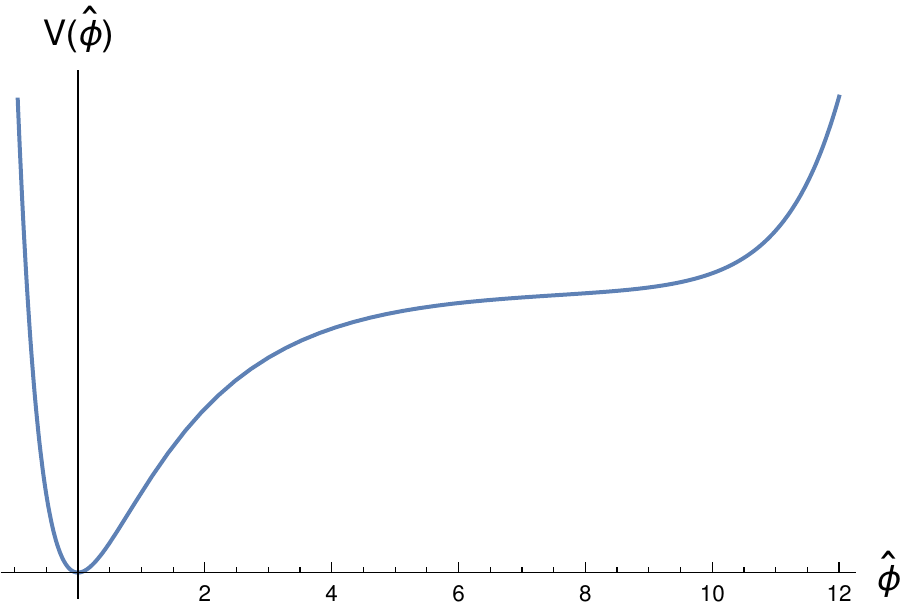}
\caption{The inflationary potential \eqref{eq:FIpot} for fibre inflation with $\lambda=10^{-6}$ features a plateau-like region very similar to Starobinsky inflation.} 
\label{fig:FibrePotential}
\end{figure}

Instead of working with $\tau_1$ and $\tau_2$, we will define two new fields, $\rho$ and $\phi$. These are the physical mass eigenstates which diagonalise the mass matrix. They are related to $\tau_1$ and $\tau_2$ through \cite{Cicoli:2021skd}
\begin{eqnarray}
\ln\tau_1&=& \sqrt{\frac{2}{3}}\,\rho + \frac{2}{\sqrt{3}}\,\phi\,, 
\label{eq:Mass_Eigenstates_and_tau} \\
\ln\tau_2&=& \sqrt{\frac{2}{3}}\,\rho - \frac{1}{\sqrt{3}}\,\phi \,.
\end{eqnarray}
Notice that $\rho$ corresponds to the volume, while $\phi$ gives the ratio $u$ between $\tau_1$ and $\tau_2$
\begin{equation}
\begin{split}
&\mathcal{V}=\sqrt{\tau_1}\tau_2=e^{\sqrt{\frac{3}{2}}\rho}\;,\\
    &u \equiv \frac{\tau_1}{\tau_2}=e^{\sqrt{3}\phi}.
\end{split}
\end{equation}
Let us consider the field $\phi$ and shift it from its minimum $\phi= \langle \phi \rangle + \hat\phi$. The potential takes the form
\begin{equation}
V(\hat\phi)=V_0 \left( 3- 4\,e^{-\hat\phi/\sqrt{3}} + 
 e^{-4\hat\phi/\sqrt{3}} + \lambda\, e^{2\hat\phi/\sqrt{3}}\right)\,,
 \label{eq:FIpot}
\end{equation}
where $V_0 \equiv \mathcal{O}(1)\times \mathcal{V}_0^{-10/3}$ is the inflaton-independent uplifting contribution and $\lambda\equiv 16 g_s^4 
 AC/B^2 \propto g_s^4 \ll 1$. This potential is shown in Fig. \ref{fig:FibrePotential} and it resembles Starobinsky inflation very closely. The similarity between the two is clearer if we focus on the slow-roll region where the fibre inflation potential can be very well approximated as
\begin{equation}
V(\hat\phi)\simeq V_0 \left( 1 - \frac{4}{3}\,e^{-\hat\phi/\sqrt{3}} \right)\;.
\label{Vsimpl}
\end{equation}
In fact, ref. \cite{Broy:2014xwa} tried to apply the standard $f(R)$ duality to fibre inflation with approximated potential (\ref{Vsimpl}) finding that its version in Jordan frame would be an $f(R)$ theory with $f(R) = R^{2-1/\sqrt{2}} + R^2 \simeq R^{1.3} + R^2$ which is very similar to the Starobinsky model $f(R)=R+R^2$. Moreover, fibre inflation predicts a relation among the scalar spectral index $n_s$ and the tensor-to-scalar ratio $r$ of the form $r=6 (n_s-1)^2$ which is almost analogous to the one of Starobinsky inflation $r=3 (n_s-1)^2$ \cite{Burgess:2016owb}.

As can be clearly seen in Fig. \ref{fig:FibrePotential}, the fibre inflation potential (\ref{eq:FIpot}) features a rising behaviour for large field values. Again this is in complete analogy with the case of Starobinsky inflation, where higher derivative corrections make the potential too steep to drive inflation at large field values, as demonstrated in Fig. \ref{fig:StaroPotCorr} for the case of $R^4$ corrections. There the coefficient $\lambda$ has to satisfy $|\lambda|\ll 1$ to preserve inflation. However, as explained in Sec. \ref{Genf(R)}, $\lambda$ is expected to be of $\mathcal{O}(1)$ from effective field theory arguments, as well as from explicit dimensional reduction of ten-dimensional higher derivative effects in string theory. In fibre inflation, on the other hand, the situation is improved since the smallness of $\lambda$ is now guaranteed by the fact that it turns out to be proportional to the string coupling, which is a priori required to be small in order to trust perturbation theory, $\lambda \propto g_s^4 \lesssim 10^{-4}$ for $g_s\lesssim 0.1$.

\subsubsection*{Couplings to matter}
\label{subsubsec:fibre_couplings}

We now proceed to compute the couplings to matter of the two moduli $\tau_1$ and $\tau_2$. Working in the large volume regime, we have
\begin{equation}
\sqrt{\tau_1} \tau_2 \gg \tau_s^{3/2}\qquad \Rightarrow \qquad 
\mathcal{V} \simeq \sqrt{\tau_1} \tau_2\,.
\end{equation}
The tree-level K\"ahler potential is given by 
\begin{equation}
K=-2\,\ln\,\mathcal{V} \simeq -\ln\,\tau_1 -2\,\ln\,\tau_2\,.
\end{equation}
The kinetic Lagrangian for the moduli becomes (at leading order)
\begin{equation}
\mathcal{L}_{\textrm{kin}}= K_{i\bar{j}} \partial_\mu T_i \partial^\mu \bar{T}_{\bar{j}}= \frac{1}{4 \tau_1^2}\, \partial_\mu \tau_1 \partial^\mu \tau_1 + \frac{1}{2\tau_2^2}\, \partial_\mu \tau_2 \partial^\mu \tau_2\,.
\end{equation}
After including matter fields, the relevant Lagrangian takes the form
 \begin{equation}
\mathcal{L}=\frac{1}{4\,\tau_1^2}\,  \partial_\mu \tau_1 \partial^\mu \tau_1  + \frac{1}{2\,\tau_2^2}\, \partial_\mu \tau_2 \partial^\mu \tau_2 + K_{\bar{\psi} \psi} i\bar{\psi} \gamma^\mu \partial_\mu \psi - e^{K/2}\,\tilde{m}\,\bar{\psi}\psi\;.
\end{equation}
Before proceeding, we have to specify the form of the K\"ahler metric for the matter fields. We start by considering the case where the Standard Model is localised in the extra dimensions since it is realised by either D7-branes wrapped around a shrinkable four-cycle $\tau_s$ or by D3-branes at singularities. As listed in (\ref{SMonS}) and (\ref{SMonD3}) for a simple Swiss-cheese geometry, in this case the matter K\"ahler metric would scale as $K_{\bar{\psi}\psi} \sim \tau_b^{-1}$. This expression can be used also in our case after substituting $\tau_b\sim \mathcal{V}^{2/3}$, and then generalising to
\begin{equation}
K_{\bar{\psi}\psi}
\simeq \frac{1}{\mathcal{V}^{2/3}}\simeq \frac{1}{\tau_1^{1/3}\tau_2^{2/3}}\,,
\end{equation}
where we ignored the dependence of the K\"ahler matter metric on $\tau_s$ since it would induce just a subdominant contribution to the couplings of $\tau_1$ and $\tau_2$. Moreover, we have
\begin{equation}
e^{K/2}\simeq \frac{1}{\sqrt{\tau_1}\,\tau_2}\;.
\end{equation}
Using these expressions, we arrive at the following Lagrangian
\begin{equation}
\mathcal{L}=\frac{1}{4\tau_1^2}\, \partial_\mu \tau_1 \partial^\mu \tau_1  + \frac{1}{2\tau_2^2}\,\partial_\mu \tau_2 \partial^\mu \tau_2 +\frac{1}{\tau_1^{1/3}\tau_2^{2/3}}\, i\bar{\psi} \gamma^\mu \partial_\mu \psi - \frac{1}{\sqrt{\tau_1}\tau_2}\,\tilde{m}\,\bar{\psi}\psi\;.
\end{equation}
We can now canonically normalise the fields as follows
\begin{equation}
\label{eq:Can_Norm_Fields}
\tau_1=e^{\sqrt{2}\phi_1}\;,\qquad \tau_2=e^{\phi_2}\;,\qquad \frac{\psi}{\tau_1^{1/6}\tau_2^{1/3}}=\psi_c\;,
\end{equation}
The Lagrangian in terms of the canonical fields looks like
\begin{equation}
\label{eq:Normalized_Moduli_Lagrangian}
\mathcal{L}=\frac{1}{2}\,\partial_\mu \phi_1 \partial^\mu \phi_1  +\frac{1}{2}\,\partial_\mu \phi_2 \partial^\mu \phi_2 +i \bar{\psi}_c\gamma^\mu \partial_\mu \psi_c - e^{-\frac{\sqrt{2}}{6}\,\phi_1}\,e^{-\frac{1}{3}\phi_2}\,\tilde{m}\,\bar{\psi}_c\psi_c\;.
\end{equation}
We can now read off the fermionic mass which is given by
\begin{equation}
m =     e^{-\frac{\sqrt{2}}{6}\,\langle\phi_1\rangle}\,e^{-\frac{1}{3}\langle\phi_2\rangle}\, \tilde{m}\,.
\end{equation}
Let us now focus on the last term of \eqref{eq:Normalized_Moduli_Lagrangian}, shift the fields as $\phi_i = \langle\phi_i\rangle + \hat\phi_i$, and expand the exponentials up to linear order to get
\begin{equation}
\label{eq:Normalized_Moduli_Lagrangian_Expanded}
\mathcal{L} \supset -\left(1-\frac{\sqrt{2}}{6}\hat\phi_1\right)\,\left(1-\frac{\hat\phi_2}{3}\right)\,m\, \bar{\psi}_c \psi_c \simeq - m\,\bar{\psi}_c \psi_c + \frac{\sqrt{2}}{6} m\,\hat\phi_1\,\bar{\psi}_c\psi_c + \frac{1}{3}\,m\,\hat\phi_2\,\bar{\psi}_c \psi_c\;.
\end{equation}

Recall that we are interested in the mass eigenstates $\rho$ and $\phi$. Out of \eqref{eq:Mass_Eigenstates_and_tau}, it is straightforward to obtain a relation between $\rho$, $\phi$ and $\phi_1$ and $\phi_2$
\begin{eqnarray}
\phi_1&=&\frac{1}{\sqrt{3}}\, \rho + \sqrt{\frac{2}{3}}\, \phi\,,
\label{eq:Mass_Eigenstates_and_phi} \\
\phi_2&=&\sqrt{\frac{2}{3}}\, \rho - \frac{1}{\sqrt{3}}\, \phi
\end{eqnarray}
Substituting these expressions into \eqref{eq:Normalized_Moduli_Lagrangian_Expanded} we get
\begin{equation}
\begin{split}
\mathcal{L} &\supset -m\, \bar{\psi}_c \psi_c +\frac{\sqrt{2}}{6}m \left( \frac{1}{\sqrt{3}}\hat\rho + \sqrt{\frac{2}{3}} \hat\phi \right)\,\bar{\psi}_c \psi_c +\frac{1}{3} m \left( \sqrt{\frac{2}{3}} \hat\rho - \frac{1}{\sqrt{3}} \hat\phi \right)\,\bar{\psi}_c \psi_c \\
&= -m\, \bar{\psi}_c \psi_c +\frac{1}{\sqrt{6}}\,m\, \hat\rho\, \bar{\psi}_c \psi_c\;,
\end{split}
\end{equation}
from which we can easily read off the couplings to matter of the physical fields
\begin{equation}
y_\rho = -\frac{1}{\sqrt{6}}\;, \qquad y_\phi = 0\;.
\end{equation}
We see that the field $\rho$ has the right Yukawa coupling to matter. This is not surprising since $\rho$ corresponds to the overall volume mode which we have already shown to feature the same coupling to matter as the Starobinsky field, even if its potential is just a runaway at large field values. 

On the contrary, $\phi$ is effectively decoupled at this level of approximation. A non-zero $y_\phi$ would be induced by any perturbative corrections to the matter K\"ahler metric which depends explicitly on the mode $u=\tau_1/\tau_2$ orthogonal to $\mathcal{V}$. However this coupling would be much weak than Planckian due to volume suppression factors. However, we have seen that a suitable scalar potential with the desired plateau can be constructed for $\phi$ (or equivalently $u$). This motivates the search for other D-brane configurations leading to different choices for the K\"ahler metric of matter fields. Since $\phi$ features a promising inflationary potential, the hope is to find a configuration for which it also couples to matter appropriately. 

Let us therefore consider the situation where the Standard Model lives on D7-branes wrapped around the two bulk cycles $\tau_1$ and $\tau_2$ which control the Calabi-Yau volume. In analogy with the Swiss-cheese case summarised in (\ref{SMonB}), we expect modular weights which can take values in $\{0,1/2,1\}$. To be as general as possible, we consider however a K\"ahler matter metric with arbitrary modular weights
\begin{equation}
K_{\bar{\psi}\psi}\simeq\frac{1}{\tau_1^{\lambda_1}\tau_2^{\lambda_2}}\,.
\label{eq:Alpha_Kahler}
\end{equation}
We determine now the values of the modular weights for which the coupling of $\phi$ to matter reproduces $y_\phi = -1/\sqrt{6}$. The starting point is the following Lagrangian
\begin{equation}
\mathcal{L}=\frac{1}{4\tau_1^2}\,  \partial_\mu \tau_1 \partial^\mu \tau_1  +\frac{1}{2\tau_2^2}\, \partial_\mu \tau_2 \partial^\mu \tau_2+\frac{1}{\tau_1^{\lambda_1}\tau_2^{\lambda_2}}\, i\bar{\psi} \gamma^\mu \partial_\mu \psi - \frac{1}{\sqrt{\tau_1}\tau_2}\,\tilde{m}\,\bar{\psi}\psi\;.
\end{equation}
The canonically normalised versions of $\tau_1$ and $\tau_2$ are still given by $\phi_1$ and $\phi_2$ as in \eqref{eq:Can_Norm_Fields}. 
The matter fields, instead, are normalised according to
\begin{equation}
\frac{\psi}{\tau_1^{\lambda_1/2}\tau_2^{\lambda_2/2}} =\psi_c\;.
\end{equation}
Having canonically normalised the fields, let us focus on the relevant term in the Lagrangian
\begin{equation}
\label{eq:Alpha_Lagrangian}
\mathcal{L} \supset -\tau_1^{(\lambda_1 -\frac{1}{2})}\tau_2^{(\lambda_2-1)}\,\tilde{m}\,\bar{\psi}_c \psi_c\;.
\end{equation}
Using 
\begin{equation}
\tau_1=e^{\sqrt{2}\phi_1}\,,\qquad \tau_2=e^{\phi_2}\,,
\end{equation}
substituting this result in the Lagrangian \eqref{eq:Alpha_Lagrangian}, and expanding to first order, we obtain the following cubic couplings
\begin{equation}
\begin{split}
\mathcal{L} &\supset -\left( 1+\sqrt{2} \left(\lambda_1 - \frac{1}{2}\right) \, \hat\phi_1 \right) 
\left( 1 + \left(\lambda_2 -1\right)\, \hat\phi_2 \right)\,\tilde{m}\, \bar{\psi}_c \psi_c \\
&\supset - \tilde{m}\,\bar{\psi}_c \psi_c 
- \left( \sqrt{2} \left(\lambda_1 - \frac{1}{2}\right) \, \hat\phi_1 + \left(\lambda_2 -1\right)\, \hat\phi_2 \right)
\tilde{m}\,\bar{\psi}_c \psi_c\;.
\end{split}
\end{equation}
At this point, we substitute the expressions \eqref{eq:Mass_Eigenstates_and_phi} giving $\phi_1$ and $\phi_2$ in terms of $\rho$ and $\phi$ and obtain
\begin{equation}
\mathcal{L} \supset - \tilde{m}\,\bar{\psi}_c \psi_c \left[ 
\sqrt{\frac23}\left(\lambda_1+\lambda_2-\frac32 \right) \hat\rho 
+ \frac{1}{\sqrt{3}}\left(2\lambda_1-\lambda_2\right)\hat\phi \right] \;.
\end{equation}
The Yukawa coupling to matter of the inflaton field $\hat\phi$ of fibre inflation models takes therefore the generic form
\begin{equation}
y_\phi =  \frac{1}{\sqrt{3}}\left(2\lambda_1 -\lambda_2\right).  
\end{equation}
This Yukawa coupling can match the conformal coupling of the Starobinsky scalar $y_\phi = -1/\sqrt{6}$ only if
\begin{equation}
2\lambda_1-\lambda_2 =-\frac{1}{\sqrt{2}}\,.
\label{eq:result}
\end{equation}
This equation can never be satisfied if the modular weights are rational numbers, implying that fibre inflation cannot reproduce the conformal coupling to matter typical of Starobinsky inflation. As already pointed out, the K\"ahler metrics for chiral matter fields on different brane setups in toroidal/orbifold orientifolds have been studied in \cite{Aparicio:2008wh}, noting that the modular weights are generally rational numbers. For the toroidal case, the only three possibilities are $\lambda_i\in\{0,1/2,1\}$. In more general setups, it is hard to determine the modular weights directly. Fortunately they normalise the physical Yukawa couplings, which are also given by the triple overlap of wave functions in the extra dimensions \cite{Conlon:2006tj}. Thereby one can extract the sum of three modular weights,\footnote{Unless the unnormalised coupling vanishes, i.e. there is no trilinear coupling between the fields.} which is related to the scaling of the wave functions with the cycle volumes given by the moduli. It is clear that there are no irrational numbers involved here, so also the individual modular weights must be rational. Hence \eqref{eq:result} does not seem to be achievable. 

For rational modular weights, the Yukawa couplings of fibre moduli to fermions are expected to be of $\mathcal{O}(1)$ with exact numerical values which depend on the D-brane origin of matter fields. As illustrative examples, we consider two cases following again \cite{Aparicio:2008wh}. Fermions at the intersection between a D7-stack wrapping $\tau_1$ and another D7-stack wrapping $\tau_2$ would be characterised by $\lambda_1=1/2$ and $\lambda_2=0$, resulting in $y_\phi = 1/\sqrt{3}$. On the other hand, open string modes living within the worldvolume of a D7-stack wrapping $\tau_1$ would have $\lambda_1=0$ and $\lambda_2=1$, resulting in $y_\phi = -1/\sqrt{3}$. Note finally that the couplings of fibre moduli to gauge bosons and Higsses on $\tau_1$, as well as to ultralight bulk axions, play a relevant role for reheating and have been computed in \cite{Cicoli:2018cgu, Cicoli:2022uqa}.

\section{Conclusions}
\label{Conclusions}

In this paper we have investigated the possibility to embed Starobinsky inflation in a UV complete theory as string theory. Previous studies \cite{Burgess:2016owb} have already shed light on the behaviour of higher curvature terms arising as the low-energy limit of ten-dimensional $\alpha'$-corrections. In particular, when the $R^2$ term competes with the standard Einstein-Hilbert action, $R^n$ terms with $n>2$ can never be neglected, and tend to spoil the flatness of the inflationary plateau. 

We instead took a different approach which exploits the fact that $R+R^2$ inflation is conformally dual to a standard Einstein gravity coupled to a scalar field with a flat exponential potential and a precise Yukawa coupling to matter fermions which are minimally coupled in Jordan frame. We tried therefore to search for string moduli which can effectively reproduce these features within the four-dimensional low-energy theory of Calabi-Yau compactifications at large volume and weak string coupling. We focused on type IIB K\"ahler moduli which naturally admit an exponential potential and tend to couple to matter with Planckian strength. 

We investigated three different classes of stringy inflaton candidates: the volume modulus, blow-up modes and fibre moduli. While the overall volume modulus generically features the correct Yukawa coupling to matter, the typical runaway behaviour of its potential does not allow for the plateau necessary for Starobinsky-like inflation. Blow-up moduli couple instead too strongly to matter fermions localised on D7-branes wrapping these four-cycles, or are effectively decoupled in other Standard Model realisations. Moreover, their potential is too shallow to reproduce Starobinsky inflation. Finally, the moduli of fibre inflation models feature a promising scalar potential which leads to cosmological predictions very similar to the ones of Starobinsky inflation, for example for the scalar spectral index and the tensor-to-scalar ratio. Moreover, fibre inflation features a nice mechanism to tame higher order corrections since they are proportional to a small suppression factor of order $g_s^4\ll 1$. However fibre moduli cannot reproduce the correct Yukawa coupling to fermions in any D-brane setup which realises the visible sector. In particular, we found that to produce the universal coupling that would allow us to identify a fibre modulus as the Starobinsky inflaton, its modular weight would need to be irrational. Given that modular weights are expected to be rational \cite{Aparicio:2008wh}, engineering the correct coupling in this way seems impossible. 

We conclude therefore that providing a UV complete justification of Starobinsky inflation remains a big challenge. Our detailed computation of moduli couplings to fermions, provides also a way to discriminate between Starobinsky inflation (setting aside the issue of UV consistency) and fibre inflation.

\section*{Acknowledgements}

We would like to thank Cliff Burgess and Fernando Quevedo for useful conversations.

\bibliographystyle{JHEP}
\bibliography{references}

\end{document}